# Imprinting of Antiferromagnetic Vortex States in NiO-Fe Nanostructures

M. Ślęzak[1*], T. Wagner[2], V. K. Bharadwaj[2], O. Gomonay[2], A. Kozioł-Rachwał[1], T. O. Menteş[3], A. Locatelli[3], M. Zając[4], D. Wilgocka-Ślęzak[5], P. Dróżdż[6], T. Ślęzak[1]

[1]AGH University of Krakow, Faculty of Physics and Applied Computer Science, Kraków, Poland

[2]Institute of Physics, Johannes Gutenberg-University Mainz, 55099 Mainz, Germany

[3]Elettra - Sincrotrone Trieste S.C.p.A., Basovizza, Trieste, Italy

[4]National Synchrotron Radiation Centre SOLARIS, Jagiellonian University, Kraków, Poland

[5]Jerzy Haber Institute of Catalysis and Surface Chemistry PAS, Kraków, Poland

[6]Institute of Physics, Maria Curie-Sklodowska University, Lublin, Poland

[*]mislezak@agh.edu.pl

**ABSTRACT**

Magnetic vortices are topological spin structures frequently found in ferromagnets, yet novel to antiferromagnets. By combining experiment and theory, we demonstrate that in a nanostructured antiferromagnetic-ferromagnetic NiO(111)-Fe(110) bilayer, a magnetic vortex is naturally stabilized by magnetostatic interactions in the ferromagnet and is imprinted onto the adjacent antiferromagnet via interface exchange coupling. We use micromagnetic simulations to construct a corresponding phase diagram of the stability of the imprinted antiferromagnetic vortex state. Our in depth analysis reveals that the interplay between interface exchange coupling and the antiferromagnet magnetic anisotropy plays a crucial role in locally reorienting the Néel vector out-of-plane in the prototypical in-plane antiferromagnet NiO and thereby stabilizing the vortices in the antiferromagnet.

**Introduction**

Antiferromagnetic (AFM) materials have recently emerged as promising candidates for next-generation spintronics applications due to their unique properties such as ultrafast dynamics, negligible stray fields, and robustness against external magnetic fields[1–3]. AFM elements are already used as passive components in spintronic devices[3,4]. The key challenge in AFM materials is to develop methods for manipulating and characterizing the AFM order parameter, the so called Néel vector, without access to any net magnetization. Detection of AFM spin textures is expedited due to advancements in experimental techniques, such as x-ray magnetic linear dichroism (XMLD)[5] and photoemission electron microscopy (PEEM)[6–10], which enable the precise characterization of the magnetic properties of AFM materials at the nanoscale.

From the manipulation perspective, particular efforts have been dedicated to create and stabilize topologically protected AFM textures such as vortices or skyrmions[1,11–14], which are promising building blocks for information storage[15,16] and unconventional computing[17,18]. One possible solution is to use antiferromagnetic-ferromagnetic (AFM-FM) systems[19], in which spin textures existing in the FM layer can be imprinted onto the AFM layer via exchange coupling at the AFM-FM interface. Previous experimental studies concerning continuous films have indeed reported the imprinting of the AFM domain structure onto the ferromagnetic (FM) magnetization[4,11,20–22]. In addition, the interface exchange-mediated transfer of a FM vortex pattern to the AFM layer has been demonstrated in patterned μm-sized discs of IrMn-NiFe or CoO(NiO)-Fe[6,7,23–26].

In spite of the above observations the following fundamental questions remain unanswered: Can the AFM vortex state be stabilized at the nanoscale? How does interface AFM-FM exchange coupling contribute to the stabilization of topologically non-trivial spin textures in AFM materials?

In this paper we provide experimental and theoretical evidence for the creation of vortex states in exchange-coupled NiO(111) on self-organized Fe(110) epitaxial nanostructures grown on a W(110) single crystal surface, see Fig. 1a. We experimentally show that the vortex states are naturally stabilized in the uncovered FM nanostructures as a result of the competition between the magnetic anisotropy, magnetostatic and exchange interactions. Subsequently, due to interface exchange coupling the FM vortex structures are imprinted onto the AFM NiO layer, which is grown onto the Fe(110) nano-template. Furthermore, our theoretical studies provide evidence of an out-of-plane Néel vector component in the center of the vortex states in the AFM NiO, which indicates the potential to locally induce out-of-plane AFM moments in the archetypal in-plane antiferromagnet NiO. We conclude that the competition between interface exchange coupling and AFM magnetic anisotropy not only governs the stability of the induced AFM vortex state, as we show in a corresponding phase diagram, but also determines its core size.

**Results**

The structure of the samples investigated in the present work is presented schematically in Fig. 1a. The main experimental results are presented in Fig. 1b-g, where the magnetic texture of both the FM and AFM components of the NiO-Fe nanostructures are imaged using photoemission electron microscopy (PEEM) combined with X-ray magnetic circular and linear dichroism (XMCD and XMLD) techniques. In Fig. 1b and 1c, we present XMCD- and XMLD-PEEM images of selected $4\mu m \times 3\mu m$ regions that show the general magnetic configuration in the NiO-Fe islands. In the XMCD image (Fig. 1b) we observe the elongated Fe nanostructures surrounded by the pseudomorphic Fe monolayer that covers the W substrate. For more details concerning the morphology and the local crystallographic structure of both nanoislands and the monolayer 'sea' and the lack of magnetic contrast in the Fe monolayer we refer the reader to the *Supplementary Material*. The nanostructures with in-plane magnetization orthogonal to the incoming radiation (wave vector **k**) do not give any circular dichroic signal and thus appear in neutral gray, similar to the monolayer areas. The white and black areas correspond to local magnetization vectors parallel or antiparallel to the beam propagation direction, respectively. Two yellow, dashed and solid rectangles in Fig. 1b mark the regions where the narrow nanostructure hosts multiple FM vortices. Such magnetic vortex structures are directly imprinted into the AFM NiO overlayer, as seen in the corresponding XMLD image presented in Fig. 1c, where the analyzed vortices are marked by green dashed and solid rectangles. In Fig. 1d-g, these regions are magnified. In these images the white arrows and black double arrows schematically depict the local spin structure in Fe and NiO sublayers, respectively. XMLD is not selective with regard to parallel or antiparallel directions of the Néel vector. Only the Néel vector axes aligned with and orthogonal to the **k** vector produce contrast. Thus, contrary to the three distinct intensity levels observed in circular dichroism from FM regions, corresponding to antiparallel, parallel and orthogonal alignment between the magnetization and photon helicity vectors, only two are observed in linear dichroism from the AFM layer. The image intensity analysis and above conclusions are directly confirmed by complementary XMCD- and XMLD-PEEM images obtained for the second, orthogonal sample orientation. We refer the reader to the *Supplementary Material* for further information.

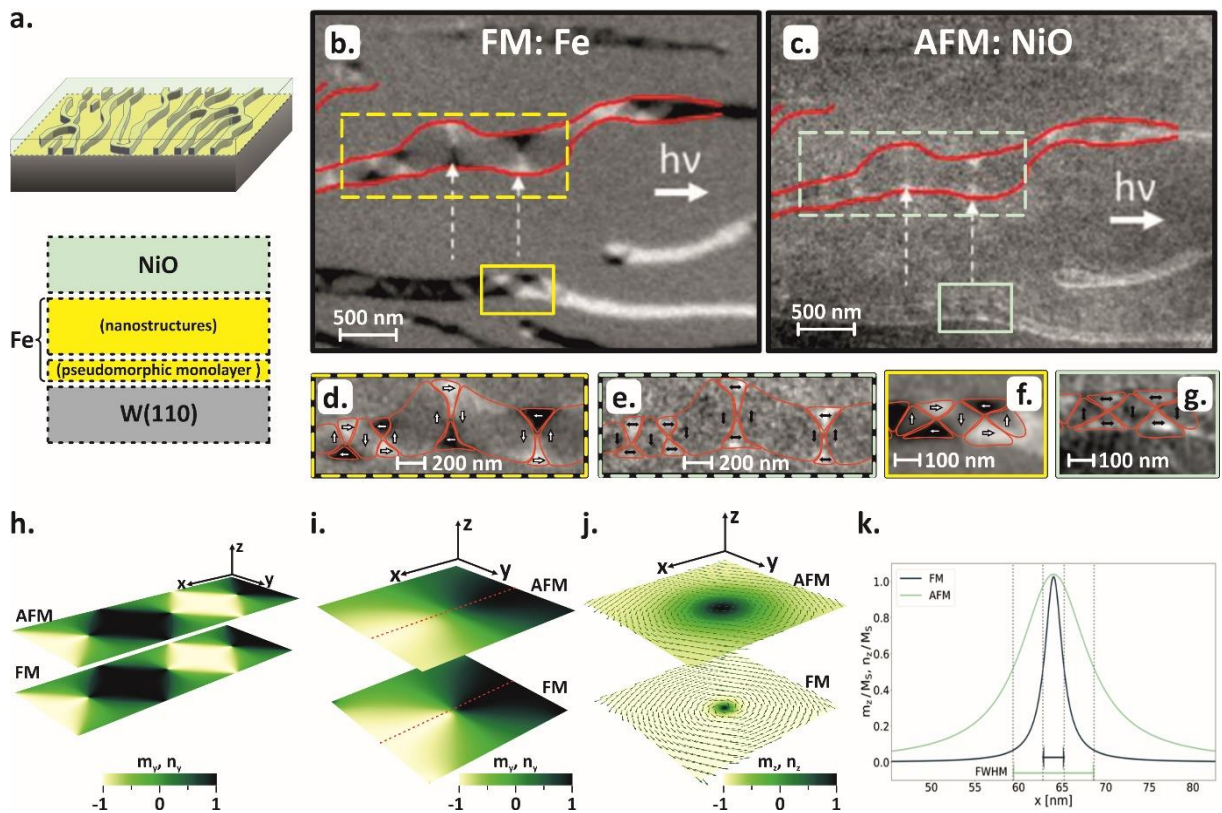

**Figure 1.** a) Schematic representation of the NiO-Fe bilayer structure. b,c) XMCD- and XMLD-PEEM images of a typical 4$\mu$m×3.0$\mu$m region covering several NiO-Fe nanostructures. Two yellow (b) and green (c) dashed and solid rectangles mark regions selected for magnification shown in d-g. White and black arrows schematically indicate the local in-plane orientation of magnetic moments in Fe and double arrows for the Néel vector for NiO vortices, respectively. h) Visualization of micromagnetic

simulation results showing three AFM-FM vortices arranged side-by-side. i, j) Magnetization (**m**) and the Néel (**n**) vector configuration in the *y* and *z* components. The upper layer represents the AFM Néel vector, while the lower layer represents the FM magnetization. The color codes indicate the magnitude of the selected order parameter. In j), the arrows indicate the in-plane components of the magnetization and of the Néel vector. k) $m_z$- and $n_z$-profiles, as depicted in i) and j). The red dashed line in (i) marks the section used for $m_z$- and $n_z$-profiles shown in (k).

While the in-plane magnetic domain structures imaged by PEEM make the identification of vortex states possible, the limited spatial resolution of 30 - 50 nm does not allow the vortex structure to be visualized in sufficient detail. According to topological arguments, to be discussed below, at the center of the vortex the magnetization and the Néel vector should point out-of-plane. To confirm the presence of out-of-plane magnetic moments and provide criteria for the generation and stability of the AFM vortex, we have performed micromagnetic simulations. We define the vortex core as the region around the center of the circular magnetization texture, where the spins point partially or fully out-of-plane. In our theoretical approach we combine both an analytical model and micromagnetic simulations, as detailed in the *Methods* section. In our model we fix the FM vortex. Next, we exchange-couple the AFM spin system to the ferromagnet at the interface, and determine the AFM ground state. Following a recent report on continuous NiO(111)-Fe(110) bilayers[9], our simulations assume parallel NiO-Fe interface exchange coupling.

We identify the AFM layer thickness as a parameter that effectively alters the relative influence of the AFM anisotropy contribution to the total energy of the system. Accordingly, we present the results of our analytical calculations of the AFM state in terms of the AFM layer thickness $t_{AFM}$ and the interface exchange coupling strength $J_{coup}$. We first fix $t_{AFM}$ = 4 nm (as used in the experiment) and focus on the strong AFM-FM interface coupling scenario. In relation to known literature on interface exchange coupling we consider $J_{coup}$ = 2x10$^{-2}$ J/m$^2$ to be strong coupling[20]. We present the corresponding simulated Fe and NiO vortex states in Fig. 1h. The computed spin structures not only reproduce well the experimental distribution of the in-plane magnetization and Néel vector components (Fig. 1f, g) but also prove the presence of the out-of-plane magnetic moments around the vortex center. Please note the alternating spin orientation for adjacent vortices that is accompanied by the opposite sense of rotation of the in-plane spin texture (Fig. 1h), consistently with experimental data (Fig. 1f, g). Importantly, with the assumed $J_{coup}$ value, we predict a perfect imprinting of the FM spin structure onto the neighbouring AFM layer, including the fully collinear alignment of the FM magnetization and AFM Néel vector throughout the sample, see Fig. 4S in the *Supplementary Material*. For a given AFM thickness, one could expect that decreasing the AFM-FM exchange coupling should result in suppression of out-of-plane AFM magnetic moments in (111) oriented NiO and final disappearance of the NiO vortex. Our micromagnetic simulations clearly contradict such a description of AFM vortex state stability as the formation of AFM vortex states in NiO strongly depends on the relation between its magnetic anisotropy and interface exchange coupling strength with Fe. In Fig. 1i, j we show the comparison of the simulated spatial distribution of magnetic moments in Fe and NiO for smaller exchange coupling strength ($J_{coup} = 6 \times 10^{-4}$ J/m$^2$) and fixed $t_{AFM} = 4.8$ nm. While the general structure of the vortex state in the AFM layer closely matches the FM vortex state, we observe that the rotation of the spins from an out-of-plane orientation at the vortex center to an in-plane orientation at the nanostructure's edge extends over a larger area in the AFM vortex as compared to the FM one. As a result, the AFM vortex core size significantly exceeds the size of its adjacent FM counterpart (Fig. 1k).

**Discussion**

In general, the absence of magnetic stray fields in non-chiral compensated antiferromagnets excludes the spontaneous formation of an isolated vortex state in an AFM layer. In fact, for the model NiO(111) system the strong out-of-plane magnetic anisotropy aligns the Néel vector in the (111) plane. However, our experimental results, along with simulations for NiO-Fe nanostructures, show that the interface exchange coupling between a FM layer hosting vortices and an AFM layer also stabilizes vortex states in the AFM layer. In particular, such an imprinting of the vortex state occurs when the AFM anisotropy contribution to the total energy of the system is overcome by the AFM-FM interface exchange coupling energy. This can be clearly seen in a phase diagram in Fig. 2a. The insets A-F in Fig. 2a show the color coded $n_z$ component of the Néel vector while its in-plane components $n_x$ and $n_y$ are marked by arrows. The angle of the Néel vector with the plane's normal vector is given by $\theta$. In the case when the AFM-FM interface exchange contribution to the energy of the system dominates over AFM magnetic anisotropy, the out-of-plane orientation of the Néel vector ($\theta = 0$) is stabilized in the center of the vortex core. For a given thickness of AFM, the lowering of the AFM-FM interaction leads to continuous increase of the stable AFM vortex core size which is well seen for example from the F-D-A insets series in Fig. 2a. Below the threshold value of the interface exchange coupling, the Néel vector reorients into the $\theta = \pi/2$ in-plane orientation, which means that AFM magnetic anisotropy dominates and AFM vortex state is no longer stable. The critical value of the interface exchange coupling depends on the thickness of the AFM layer, as can be seen from the color code in Fig. 2a. Hence, we find that the imprinting is mainly determined by the competition of magnetic anisotropy, weighted by the AFM layer thickness, and the interface exchange coupling.

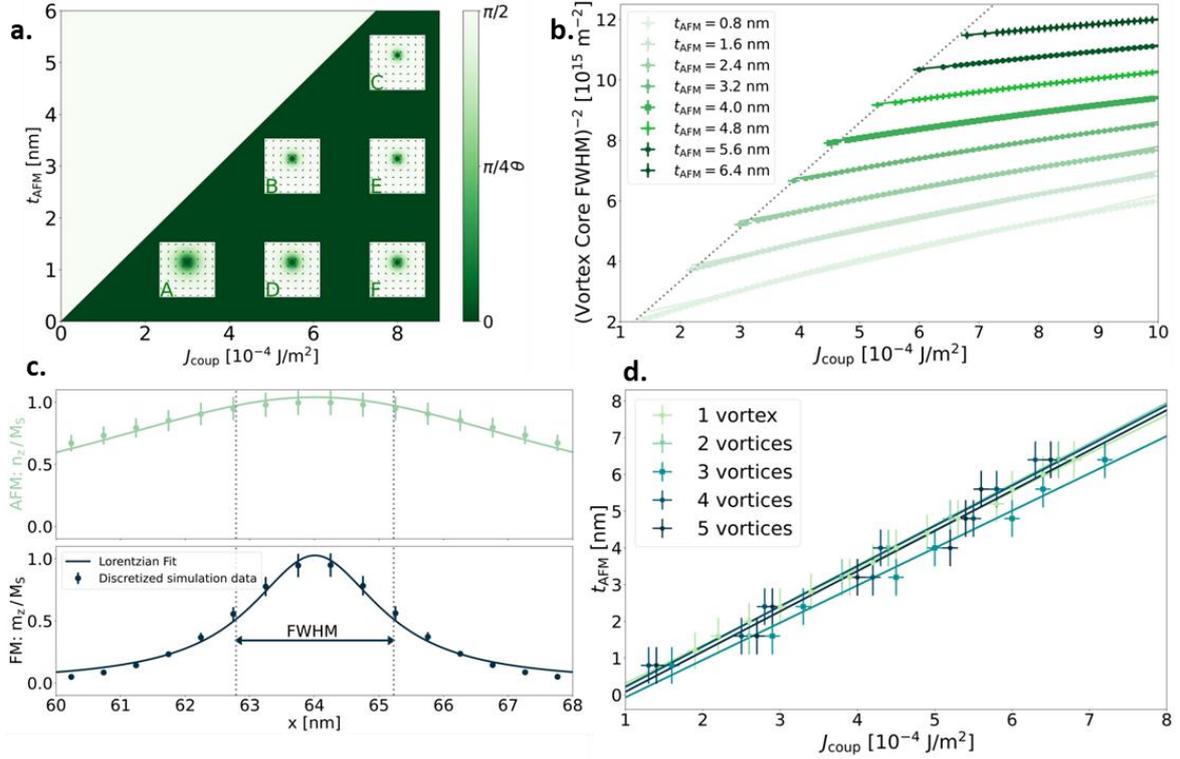

**Figure 2** a) Phase diagram of the AFM magnetic texture. Insets A-F show the $n_z$-component as color and $n_x$ and $n_y$ as arrows, obtained by micromagnetic simulations. Their position on the phase diagram is qualitative. b) Variation of vortex center full width half maximum (FWHM) of the Lorentzian fit to $n_z$ (lines) as a function of $J_{coup}$ for different AFM layer thicknesses (data points). The dotted gray line indicates the phase boundary, as also observed in panel a. We show the untransformed data in Fig. 5S. c) FM $m_z$ and AFM $n_z$ profiles through the nanoparticle center line along the x-direction. Data points show the discretized micromagnetic results, lines show Lorentzian fits to the data. The AFM vortex core FWHM is larger than the FM one. d) Micromagnetic (points) phase diagram for the AFM magnetic texture for one to five vortices in both layers. We fit the data with a linear fit (lines). We refer to the *Supplementary Material* for more details on all panels.

The effective magnetic anisotropy $K_{an}$ of the AFM given by the analytical expression $K_{an} \equiv J_{coup}/t_{AFM} - K_{AFM}$ is composed of two contributions, namely the out-of-plane AFM anisotropy $K_{AFM}$ and interface exchange coupling $J_{coup}$. We fit the numerical data according to $K_{an} = J_{coup}/t_{AFM} - K_{AFM}$ in Fig. 2b (see the *Supplementary Material* for more details). Increasing either the AFM layer thickness or the interface exchange coupling strength leads to a smaller vortex size. In the FM, the magnetostatic interactions from the sample's surface introduce a strong out-of-plane anisotropy that favors the magnetization to lie in the surface plane. Additionally, the magnetostatic interactions from the boundaries of the nanoparticles stabilize the vortex texture. However, in the AFM there are no magnetostatic interactions to stabilize the vortex texture. Instead, the interface exchange coupling to the FM magnetization leads to an imprinting of the vortex state.

While the interface exchange coupling stabilizes the AFM vortex, the anisotropy can significantly influence the size of the vortex core. The size of an AFM vortex core can, depending on the material parameters used in simulation, significantly exceed the size of adjacent FM vortex core. Specifically, for a 4 nm thick AFM layer (as used in experiment) and an AFM-FM exchange coupling of $J_{coup} = 6 \times 10^{-4}$ J/m$^2$ (central to the stable imprinted AFM vortex phase, cf. Fig. 2b) we predict 2 nm (FM) and 9 nm (AFM) wide vortex cores as presented in Fig. 2c, respectively. Such additional stabilization of the out-of-plane Néel vector component in the vicinity of vortex core is at a first glance surprising for an archetypal easy-plane antiferromagnet. Moreover, a non-trivial, locally non-collinear magnetic configuration forms at the AFM-FM interface. For example in Fig. 2c the AFM vortex core is 4.5-times as large as the FM vortex core. Such a phenomenon can be understood by taking into account the internal exchange stiffness of the AFM material. The distribution of the AFM Néel vector and the FM magnetization within the vortex structure can be considered as a partial domain wall connecting the in-plane orientation at the sample's edge with the out-of-plane orientation at the vortex center. The width of such a domain wall, denoted $x_{DW}$, scales as $x_{DW} \propto \sqrt{J_{AFM}/K_{an}}$, where $J_{AFM}$ is the exchange stiffness of the AFM material and the $K_{an}$ is the AFM effective magnetic anisotropy. Thus, the interplay between AFM-FM interface coupling, exchange stiffness and magnetic anisotropy in both the AFM and the FM materials individually determines the vortex core size in each material. The above considerations are valid only when the interface coupling $J_{coup}$ exceeds its critical value required for AFM vortex formation, i.e. within the bottom-right regions of the phase diagrams in Fig. 2a, b.

In Fig. 2d we also present the phase diagram of the stability of up to five AFM vortices chains. Again, the critical value of $J_{coup}$ required for the stabilization of imprinted vortex states in the antiferromagnet scales linearly with the thickness of the AFM layer. From the results of micromagnetic simulations presented in Fig. 2d we conclude that also multiple vortex states naturally arise in the ferromagnet and can be imprinted onto the AFM layer. The multiple vortices align side-by-side in the nanostructures and the direction of the out-of-plane magnetization at the vortex center alternates from vortex to vortex. This is in accordance with the conditions posed by topology that give rise to an alternating sense of rotation for adjacent vortices. The simulations in Fig. 2d are in agreement with already presented results in Fig. 1h on the three-vortex state and also agree with the experimental observations shown in Figs. 1b-g. Thus, the stability of the imprinted AFM vortex state is not affected by the number of vortices that are transferred from FM component.

**Conclusions**

In conclusion, we demonstrate the stabilization of topological vortex textures in the in-plane antiferromagnetic NiO(111) at zero external magnetic field. By selectively probing the nanoscale spin orientations of individual epitaxial AFM-FM nanostructures in the NiO(111) (4nm)-Fe(110) (6nm) bilayer system using XMLD- and XMCD-PEEM, we observe that the FM vortex states are successfully imprinted into the AFM spin texture.

Our analytical calculations reveal that interface exchange coupling between the AFM and the FM layers induces a local reorientation of the Néel vector to be aligned with the FM magnetization at the vortex center. Our micromagnetic simulations demonstrate that the stability of such vortex states can be tuned by the AFM layer thickness. We construct a phase diagram that takes into account the interplay between interface exchange coupling and the AFM magnetic anisotropy. The phase diagram remains consistent regardless of the number of imprinted vortices.

We find that the minimal interface exchange coupling strength required to transition into the stable AFM vortex state phase increases with AFM layer thickness. Specifically, for an AFM layer thickness of 4 nm, as studied in experiment, we calculate the critical interface exchange coupling $J_{coup} \gtrsim 6 \cdot 10^{-4} J/m^2$. Moreover, depending on the magnetic anisotropy and interface exchange coupling strength we predict possible significant differences between the AFM and FM vortex core sizes. Our results indicate the possibility of inducing localized out-of-(111)-plane AFM moments in the well-known in-plane AFM NiO(111) system.

## Methods

**Preparation and structural characterization of the samples**

The continuous, eight monolayers (ML) thick Fe(110) film was grown by molecular beam epitaxy (MBE) on an atomically clean W(110) single crystal at room temperature and subsequently post annealed at 800 K, leading to the formation of nanostructures with lateral dimensions of the order of several hundreds of nanometers and an average height estimated to be ~ 6 nm. The areas between these self-organized islands are covered with a pseudomorphic monolayer of Fe, hereafter called Fe wetting layer. The whole sample area was covered by homogenous 40 Å thick NiO overlayer prepared at room temperature by reactive deposition of Ni in a partial oxygen pressure $1 \times 10^{-6}$ mbar. The local structure of selected as prepared NiO covered nanostructures was assessed using micro-low energy electron diffraction (μLEED), a special method available in LEEM, which allows to identify the local crystal structure of microscopic surface areas. Corresponding results are presented in the *Supplementary Material*.

**Magnetic characterization of the samples**

X-PEEM images presented in Fig. 1b-g of the article were performed at 120 K, in the spectroscopic photoemission and low energy electron microscope (SPELEEM) which is the end station of the nanospectroscopy beamline in Elettra synchrotron (Trieste, Italy)[29]. In the SPELEEM setup the X-rays were incident on the sample at a 16° grazing angle from the surface. Therefore, only one of the two linear polarization states was within the sample plane giving sensitivity to the change in the in-plane spin orientation of NiO. The XMLD-PEEM images of NiO magnetic domain structures shown in Fig.

1b-g were obtained by subtracting and normalizing the two PEEM images, acquired at 868.9 and 870.2 eV photon energy, which correspond to the two absorption peaks in the Ni $L_2$ edge visible in the XMLD spectra. An exemplary µXAS spectrum collected from single nanostructure area is presented in the *Supplementary Material*. The XMCD imaging was performed at the $L_3$ absorption edge of Fe and with circular X-ray polarization. Due to grazing incidence geometry of the X-PEEM end station the XMCD-PEEM technique was mostly sensitive to in-plane magnetization along the beam direction. The XMCD algebra resulted in dark contrast for beam propagation parallel to the magnetization direction, and bright contrast for the antiparallel configuration. Detailed discussion of both XMCD- and XMLD-PEEM magnetic contrast is included in the *Supplementary Material*, where results for two complementary sample beam configurations are analyzed.

**Analytical Model**

The energy of the AFM layer can be expressed as:

$$E_{AFM} = \left[\frac{1}{2} J_{AFM} (\nabla \boldsymbol{n})^2 + \frac{1}{2} K_{AFM} n_z^2\right] t_{AFM} - J_{coup} \boldsymbol{m}_F \cdot \boldsymbol{n},$$

where $J_{AFM}$ is the magnetic stiffness, $K_{AFM}$ is the AFM out-of-plane anisotropy, $t_{AFM}$ is the AFM layer thickness, **n** is the AFM Néel vector and distribution of $\boldsymbol{m}_F$ is given by the structure of the underlying FM vortex. If we assume the radial symmetry of a FM vortex, then we can reduce the problem to quasi 1D equations for the distribution of the Néel vector obtained by minimization of $E_{AFM}[\theta]$:

$$-x_{DW}^2 \frac{1}{\rho} \frac{\partial}{\partial \rho}\left(\rho \frac{\partial \theta}{\partial \rho}\right) + \left(-1 + \frac{x_0^2}{\rho^2}\right) \sin\theta \cos\theta + h_{eff} \sin[\theta - \theta_{FM}(\rho)] = 0,$$

where $x_{DW} \equiv \sqrt{A_{AFM}/K_{AFM}}$ is the thickness of the domain wall in an isolated AFM layer, and

$$h_{eff} \equiv \frac{J_{coup}}{K_{AFM} t_{AFM}},$$

is an effective magnetic field in dimensionless units induced by the interface exchange coupling with the underlying FM layer. Here θ is the Néel vector polar angle and ρ is the radial coordinate of the vortex structure. $\theta_{FM}$ is the FM polar angle, that is fixed.

**Micromagnetic Simulations**

We model the system with the open source micromagnetic software Mumax$^3$, see references [3S,4S] in *Supplementary Material*, as a bilayer system of dimensions x × 128 nm × 2 nm, where each layer has thickness 1 nm and the system length x ∝ 128 nm × n scales with the number of vortices n. The system was discretized with a mesh size of $1 \times 1 \times 1$ nm$^3$ for simulations sweeping the phase space. For simulations investigating the vortex size and profile, a mesh size of $0.5 \times 0.5 \times 0.5$ nm$^3$ was used. We choose open boundary conditions, and for the bottom, FM, layer choose to include the exchange interaction energy (exchange stiffness $J_{FM}$), demagnetization energy and easy-plane uniaxial anisotropy (anisotropy constant $K_{FM}$). For the upper, AFM, layer we include the exchange interaction energy (exchange stiffness $A_{AFM}$) and out-of-plane anisotropy (anisotropy constant $K_{AFM}$. Please note, that reference [5S] of the *Supplementary Material* uses a different prefactor of the anisotropy term and Mumax³ uses a different sign of the anisotropy constant compared to our implementation.). The two layers are coupled by interface exchange coupling between the FM and the AFM layer of the form $J_{coup}$ **m** · **n**, where $J_{coup}$ is the interface exchange coupling, **m** is the FM magnetization and **n** is the AFM Néel vector. The material and simulation parameters are given in Tab. 1. In analogy to the experiment, we simulate the transition of the AFM layer from a homogeneously in-plane initial state into a perfect imprinting of the FM layer vortex state in two steps.

In the first step, we induce a vortex state into the FM layer. The magnetization is initialized as shown as an example for a three vortex sample on Fig. 6S in the *Supplementary Material*. Then we relax the system into a metastable vortex state as shown in Fig. 6S. For suitably chosen material constants, cf. Tab. 1, the vortex state with an out-of-plane core emerges. In the case of multiple vortices the cores have alternating out-of-plane magnetization.

In the second step we add the AFM layer on top of the FM layer. Since we investigate only the static behavior of a two sublattice antiferromagnet, we can equate the direction of the magnetization as the direction of the Néel vector. We freeze the FM layer in its vortex configuration. The AFM initial state is oriented along +x (in-plane). We relax the system, meaning the AFM Néel vector, as the FM layer is fixed in place. For strong interface exchange coupling an imprinted vortex state appears in the AFM layer.

We iteratively map the phase space spanned by the AFM layer thickness $t_{AFM}$ and the interface exchange coupling $J_{coup}$. For each parameter configuration we relax the system and check for a transition of the AFM final state from the initial in-plane orientation into a perfect imprinting of the FM state.

| Boundary Conditions | open |
|---|---|
| Damping | 0.01 |
| Demagnetization | FM layer only |
| Saturation Magnetization (both AFM and FM) | $5.00 \cdot 10^6$ A/m [5S] |
| Exchange Stiffness (both $J_{AFM}$ and $J_{FM}$) | $2.10 \cdot 10^{-11}$ J/m [5S,6S] |
| FM out-of-plane anisotropy $K_{FM}$ | $1.25 \cdot 10^5$ J/m³ [7S] |
| AFM out-of-plane anisotropy $K_{AFM}$ | $5.00 \cdot 10^5$ J/m³ for $t_{AFM}$ = 4 nm [5S] |

**Tab. 1:** Values of the micromagnetic parameters and other conditions utilized in the simulations. Please note that the sign of the anisotropy in Mumax³ is different from what is used in our model. Furthermore, please consider the different prefactors of the anisotropy terms in [5S] and our implementation.

**Acknowledgements:**

This work was supported by the National Science Centre, Poland, under Grant No. 2021/41/B/ST5/01149.

A.K.-R. acknowledges Grant No. 2020/38/E/ST3/00086 founded by the National Science Centre, Poland.

Research project was partly supported by program „Excellence initiative – research university" for the AGH University of Science and Technology".

The authors acknowledge the CERIC-ERIC Consortium for the access to experimental facilities and financial support.

The research leading to this result has been supported by the project CALIPSOplus under Grant Agreement 730872 from the EU Framework Programme for Research and Innovation HORIZON 2020.

Funded by the Deutsche Forschungsgemeinschaft (DFG, German Research Foundation) - TRR 173 – 268565370 (project B15).

Supported by the Deutsche Forschungsgemeinschaft (DFG, German Research Foundation) Skyrmion SPP 2137 (Project 403233384).

**Supplementary Material**

**Morphology and Local Crystallographic Structure of NiO-Fe Nanostructures**

The LEEM image in Fig. 1Sa provides a typical example for the morphology of Fe(110) nanostructures. The μLEED pattern of the uncovered Fe(110) surface is presented in Fig. 1Sb for the Fe wetting layer surrounding all Fe nanostructures, as schematically marked by a yellow circle in Fig. 1Sa. This diffraction pattern confirms the pseudomorphic structure of the Fe monolayer on W(110), which results from the in-plane lattice spacing $a_{001}$ along the Fe[001], fitting almost perfectly the $a_{001}$ = 3.16 Å value of tungsten bulk. Next, a diffraction pattern, cf. Fig. 1Sc, was acquired from an area comprising both the wetting layer and the chosen nanostructure, see the red circle in Fig. 1Sa. The superposition of two distinct (110) diffraction patterns is clearly seen in Fig. 1Sc. The outer diffraction spots correspond to a large distance in the reciprocal space and originate from electrons backscattered by the Fe(110) nanostructure, that has a small lattice constant in real space. The inner diffraction spots, which have a small distance in the reciprocal space correspond to diffraction from Fe pseudomorphic layer with a large lattice constant in real space. Diffraction spots from Fe nanostructure indicate a smooth unreconstructed (110) surface with the in-plane lattice spacing $a_{001}$ along the Fe[001] direction $a_{001}$ = 2.90 ± 0.02 Å, which corresponds to an almost fully relaxed Fe(110)-W film when compared to 2.86 Å of bulk iron. This value, referenced to complementary thickness dependent LEED studies on continuous Fe-W(110) films as well as to the literature, indicates that the height of the selected nanostructure is at least 6 nm. The μLEED pattern from the surface of the same nanostructure covered by 40 Å thick NiO overlayer, shown in Fig. 1d, indicates hexagonal NiO(111) surface structure. Based on the observed μLEED patterns, we thus conclude that the Fe[1-10] and Fe[001] in-plane directions are parallel to NiO[-211] and NiO[01-1] directions within Fe(110)∥NiO(111) plane, respectively.

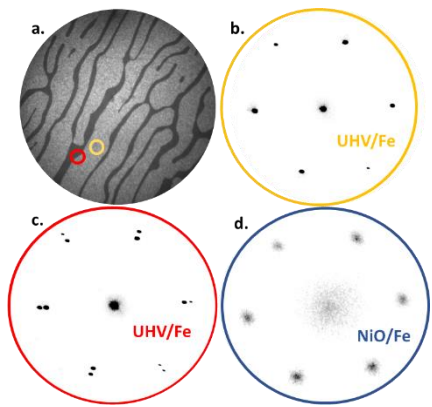

**Fig. 1S:** (a) LEEM image of NiO covered Fe(110) nanostructures formed upon annealing of 8 ML-thick Fe film on W(110) single crystal at 800 K. Field of view is 10 μm.. (b-d) μLEED patterns obtained on (b) Fe pseudomorphic monolayer surrounding the nanostructures (c) a surface area comprising both the Fe pseudomorphic monolayer and a single Fe nanostructure and (d) the same surface as area shown in (b) after growth of a homogenous NiO overlayer. Yellow and red circles in (a) schematically mark the electron beam spots for corresponding μLEED patterns presented in (b) and (c). Energies of electrons in (b,c) and (d) were 60 eV and 48 eV, respectively.

**Details of the Micro-Spectroscopy and Spectro-Microscopy Methodology**

In this section we describe XMC(L)D-PEEM micro-spectroscopy and spectro-microscopy methodology and results obtained using the spectroscopic photoemission and low energy electron microscope (SPELEEM) operating at the nano spectroscopy beamline of the Elettra synchrotron (Trieste, Italy)[1S]. In Fig. 2S a XMCD-PEEM collected at the Fe $L_3$ absorption edge is shown. The photon energy chosen for imaging is marked by a red arrow in Fig. 2Sb, where µ-XAS spectra collected with left- and right-handed (LHP and RHP) circularly polarized x-ray beam are shown, as determined from two selected domains within the nanostructure. The local orientation of Fe magnetization in these two domains is schematically marked by white and yellow arrows in Fig. 2Sa and b. The corresponding Ni $L_2$ XMLD-PEEM image of the same sample area is shown in Fig. 2Sc. The presented differential XMLD-PEEM image was obtained by subtracting the two PEEM images and normalizing to their sum, acquired at 868.9 and 870.2 eV photon energy, which correspond to the two absorption peaks in the Ni $L_2$ edge visible and schematically marked by red arrows in the µ-XAS spectrum in Fig. 2Sd.

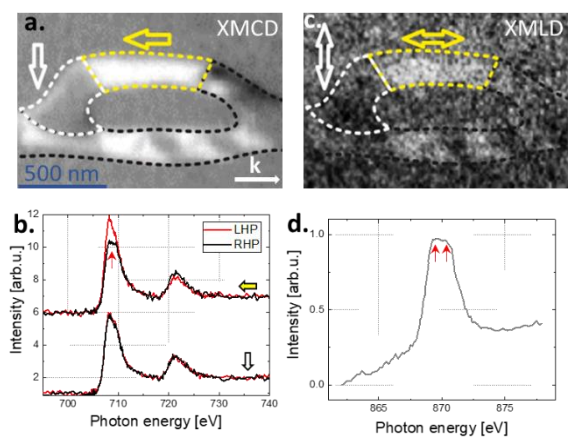

**Fig. 2S:** Micro-spectroscopy and spectro-microscopy methodology. In (a) the magnetic domain pattern of FM component in selected NiO(111)-Fe(110) nanostructure is shown as imaged by XMCD-PEEM at the Fe $L_3$ absorption edge. The photon energy chosen for imaging is marked by red arrow in (b) where µ-XAS spectra collected with left- and right-handed (LHP and RHP) circularly polarized x-ray beam are shown, as determined from two selected domains within the nanostructure. The local orientation of Fe magnetization in these two domains is schematically marked by white and yellow arrows in both (a) and (b). In (c) the corresponding Ni $L_2$ XMLD-PEEM image of the same area as in panel (a) is shown. The domain structure of antiferromagnetic NiO was imaged with the linear polarization of incoming x-rays parallel to the Fe(110) || NiO(111) surface plane. The data was collected at T ~ 120 K.

The same methodology was applied to probe an area of the sample exhibiting ferromagnetic vortex states in Fe nanostructures. Here, we acquired XMCD- and XMLD-PEEM images for two orthogonal sample orientations, in order to allow a full mapping of the magnetization in the vortices. Fig. 3Sa shows a XMCD-PEEM image acquired with the photon helicity vector **k** parallel to Fe[001] in-plane direction, providing strong magnetic contrast in the regions where the magnetization vector is parallel or antiparallel to the beam propagation direction. The corresponding XMLD image, cf. Fig. 3Sc, was acquired with the linear polarization of incoming X-rays oriented in the Fe(110)∥NiO(111) surface plane along the Fe[1-10]∥NiO[-211] direction. Fig. 3Sb and d show XMCD- and XMLD-PEEM images of the same sample area after rotation of the sample by 90°. Here, **k** is parallel to the Fe[1-10] in-plane direction and the electric field vector E with Fe[001]∥NiO[01−1] direction, respectively. Note that the XMCD images in figures 3Sa and b are complementary to each other: the nanostructure regions magnetized along in-plane axis orthogonal to incoming radiation provide intermediate, grey magnetic contrast in Fig. 3Sa while after sample rotation in Fig. 3Sb they are visible as black and white depending on the local orientation of particular magnetic domain. Please note, that for both analysed X-PEEM geometries the pseudomorphic monolayer areas surrounding Fe nanostructures appear grey in XMCD-PEEM images (Fig. 3S a,b) which means a lack of magnetic contrast. This can be attributed either to oxidation of Fe which may lead to fully oxidised monolayer or to insufficient X-PEEM sensitivity that does not allow to probe the magnetism of deeply buried single atomic layer of Fe.

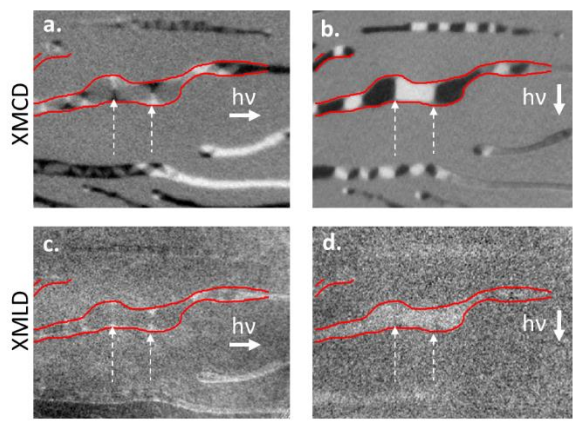

**Fig. 3S:** XCMD- (a and b) and XMLD-PEEM (c and d) images of the selected sample region obtained for two orthogonal sample orientations with respect to incoming synchrotron radiation. White dashed

arrows indicate positions of two selected magnetic vortices within central nanostructure. The field of view in all images is ~ 4 μm x 3 μm. The data was collected at T ~ 120 K.

**Supplementary Material for Fig. 1**

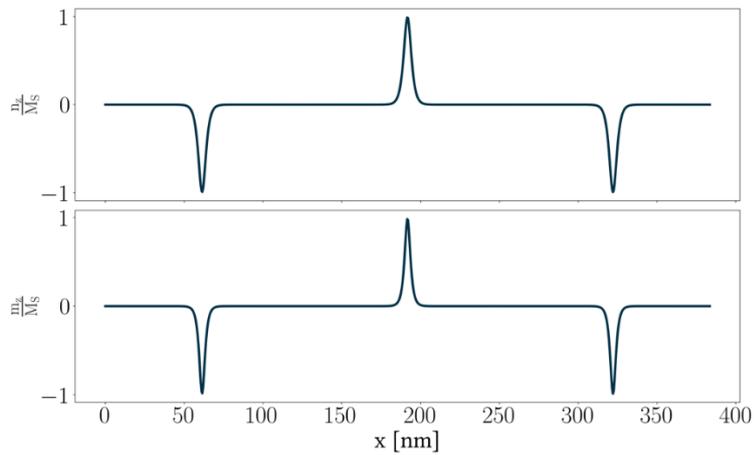

**Fig. 4S:** Perfect imprinting of the three vortex case.

**Supplementary Material for Fig. 2**

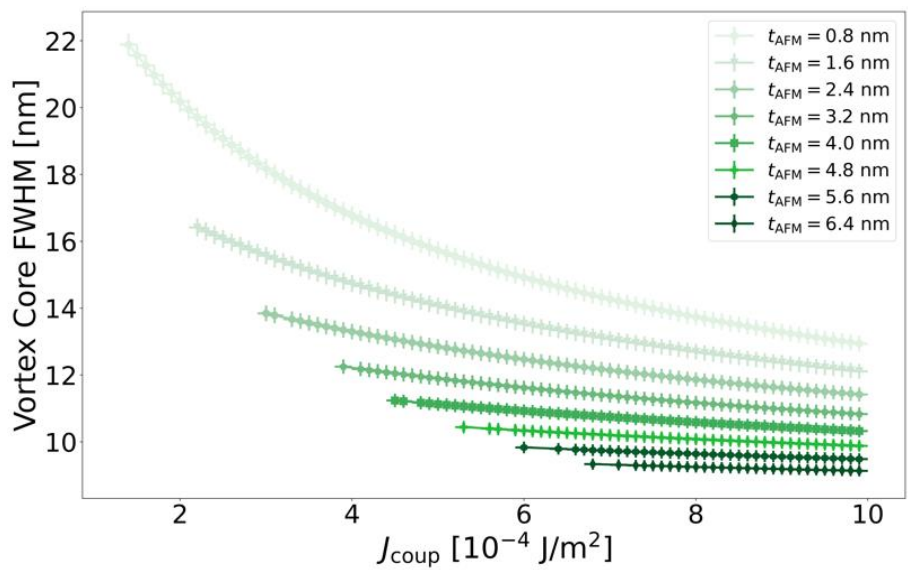

**Fig. 5S:** Vortex core scaling. In Fig. 2b we show data that has been transformed, such that a linear fit could be used. Here we show the original, untransformed data. The vortex center size in the AFM layer decreases with greater interface exchange coupling and greater AFM layer thickness.

| Picture Inset | $J_{coup}$ [$10^{-4}$ J/m$^2$] | $t_{AFM}$ [nm] |
| --- | --- | --- |
| A | 1.4 | 0.8 |
| B | 5.0 | 4.0 |
| C | 9.9 | 6.4 |
| D | 5.0 | 0.8 |
| E | 9.9 | 4.0 |
| F | 9.9 | 0.8 |

**Tab. 1:** In Fig. 2a A-F we show visualizations of the imprinted AFM vortex state from micromagnetic simulations. The position of the insets is qualitative with respect to the axes of the plot. Here we present the corresponding parameters that were used for each simulation.

| Figure | $t_{AFM}$ [nm] | $J_{coup}$ [$10^{-4}$ J/m$^2$] |
| --- | --- | --- |
| 1h | 4.0 | 200.0 |
| 1i | 4.8 | 6.000 |
| 1j | 4.8 | 6.000 |
| 1k | 4.8 | 6.000 |
| 2c | 4.8 | 6.000 |
| XS | 4.0 | 200.0 |

**Tab. 3:** Simulation parameters for visualizations.

**Comments on Solving the Analytical Model**

Close to the vortex center ($\rho \ll x_{DW}$), where $\theta_{FM} \to 0$, $\theta_{AFM} \ll 1$. In this case the solution of our analytical model can be approximated as $\theta_{AFM} = \frac{1-h_{eff}}{x_{DW}^2}\rho^2$ [2S]. By comparing with the solutions for the vortices we conclude that $K_{an} \equiv J_{coup}/t_{AFM} - K_{AFM}$ is an effective anisotropy that defines the size of an AFM vortex. Far from the vortex center ($\rho \gg x_0$), where $\theta_F \to \pi/2$, the above equation can be approximated as $(\sin\theta_{AF} + h_{eff})\cos\theta_{AF} = 0$, from which we get $\theta_{AF} = \pi/2$. Hence, the Néel vector lies in the plane.

**Micromagnetic Simulations and Data Postprocessing**

We define a perfect imprinting of the FM vortex state into the AFM state by two criteria: We check the topological charge of the AFM final state by applying

$$q = \frac{1}{4\pi}\int \mathbf{N}\cdot\left(\frac{\partial \mathbf{N}}{\partial x}\times\frac{\partial \mathbf{N}}{\partial y}\right)dx\,dy$$

where **N** is the Néel vector field and q is the topological charge. For a single vortex we detect a topological charge of 1/2, and multiples of 1/2 for multiple vortices in each layer, respectively. As a second criterion, the sign of the vortex core magnetization should be equal in the FM and the AFM layers.

We investigate the vortex center size as a function of the interface coupling. For this, we use a decreased mesh size of $0.5 \times 0.5 \times 0.5$ nm$^3$ in our simulations to increase the spatial accuracy of the magnetization and Néel vector profile. We fit a Lorentzian function to the z-component of the magnetization profile, like shown in Fig. 1h, and take the FWHM as an indicator of the relative vortex size.

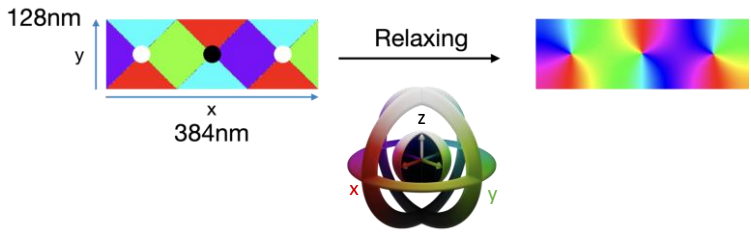

**Fig. 6S:** FM layer in its initial state and after relaxation in its final state for the case of three vortices. The system dimensions are as indicated 384 nm × 128 nm × 2 nm. An inset indicates the color code for the magnetization direction (rendering by Matthias Greber, adapted from [8S]).

**References for Supplemental material**: